\documentclass[preprint, 12pt]{elsarticle}
\usepackage[utf8]{inputenc}
\usepackage{color}
\usepackage{graphicx}
\usepackage{placeins}
\usepackage[margin=1in]{geometry}
\usepackage{appendix}
\usepackage{amsmath, amssymb, bm}
\usepackage{multirow}
\usepackage{placeins}
\usepackage{xfrac}
\usepackage[colorlinks=true, linkcolor=blue, citecolor=blue]{hyperref}
\usepackage[nameinlink, capitalize]{cleveref}
\usepackage[font=footnotesize, labelfont=bf]{caption}
\usepackage{subcaption}
\usepackage[version=4]{mhchem}
\usepackage{lineno}
\usepackage{bm}
\usepackage{comment}
\usepackage[T1]{fontenc}
\usepackage{anyfontsize}

\usepackage{tablefootnote}
\usepackage{float}
\usepackage{algorithm}
\usepackage{algorithmicx}
\usepackage{algpseudocode}

\journal{Computational Materials Science}

\begin{document}

\begin{frontmatter}

\title{Ab initio molecular dynamics of paramagnetic uranium mononitride (UN) using disordered local moments}

\author[ncsu]{Mohamed AbdulHameed}
\author[ncsu,inl]{Benjamin Beeler\corref{1}}
\cortext[1]{Corresponding author}
\ead{bwbeeler@ncsu.edu}

\address[ncsu]{Department of Nuclear Engineering, North Carolina State University, Raleigh, NC 27695}
\address[inl]{Idaho National Laboratory, Idaho Falls, ID 83415}

\begin{abstract}

This work presents an investigation of the thermophysical properties of paramagnetic uranium mononitride (UN) using \textit{ab initio} molecular dynamics (AIMD) simulations combined with the disordered local moment (DLM) approach. This methodology accurately captures the high-temperature paramagnetic state of UN, addressing the limitations of standard density functional theory (DFT) models. The AIMD+DLM model consistently predicts a cubic crystal structure for UN across all considered temperatures, aligning with experimental observations of its paramagnetic phase. Key thermophysical properties, including the lattice parameter and specific heat capacity, are computed and compared to experimental data. The calculated lattice parameter is somewhat underestimated relative to the empirical correlation, consistent with prior studies modeling UN as a ferromagnetic (FM) or antiferromagnetic (AFM) material. The specific heat capacity exhibits slight underestimation at high temperatures, while closely following the experimental trend. These results highlight the utility of the AIMD+DLM framework in modeling paramagnetic materials. A script implementing the AIMD+DLM methodology for compounds and metals using the VASP code is also provided. This tool facilitates the systematic application of the method and is expected to broaden its adoption within the computational materials science community.

\end{abstract}


\end{frontmatter}

\newpage

\section*{Highlights}
\begin{itemize}
\item AIMD+DLM enables accurate modeling of the paramagnetic state of materials
\item AIMD+DLM predicts cubic UN structure across all considered temperatures
\item Thermophysical properties agree with data, with minor underestimation
\item AFM UN is unstable and relaxes to FM UN unless enforced
\item Supplementary Bash/Python script automates AIMD+DLM setup in VASP
\end{itemize}

\section*{Graphical Abstract}
\begin{center}
\includegraphics[width=\textwidth]{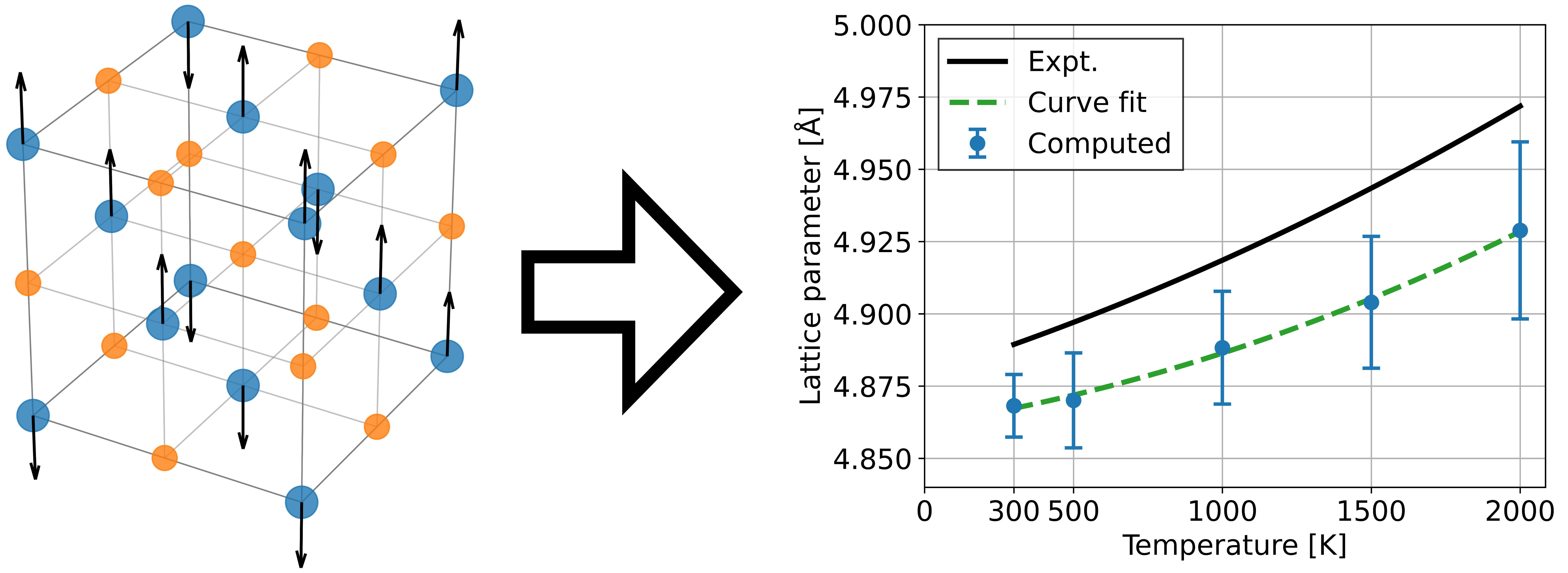}
\end{center}

\newpage

% \linenumbers

\section{Introduction}

Uranium mononitride (UN) is recognized as a promising advanced nuclear fuel, offering several advantages including high fissile density, excellent thermal conductivity, compatibility with various cladding materials, and potential for extended fuel cycles \cite{Wallenius2020, Uno2020}. Despite these benefits, certain properties of UN remain unexplored, particularly the influence of its magnetic state on various material characteristics. This is due to the challenges associated with accurately simulating its magnetic states using density functional theory (DFT).

% Experimental magnetic state of UN

\begin{figure}[h]
    \centering
    \includegraphics[width=0.5\linewidth]{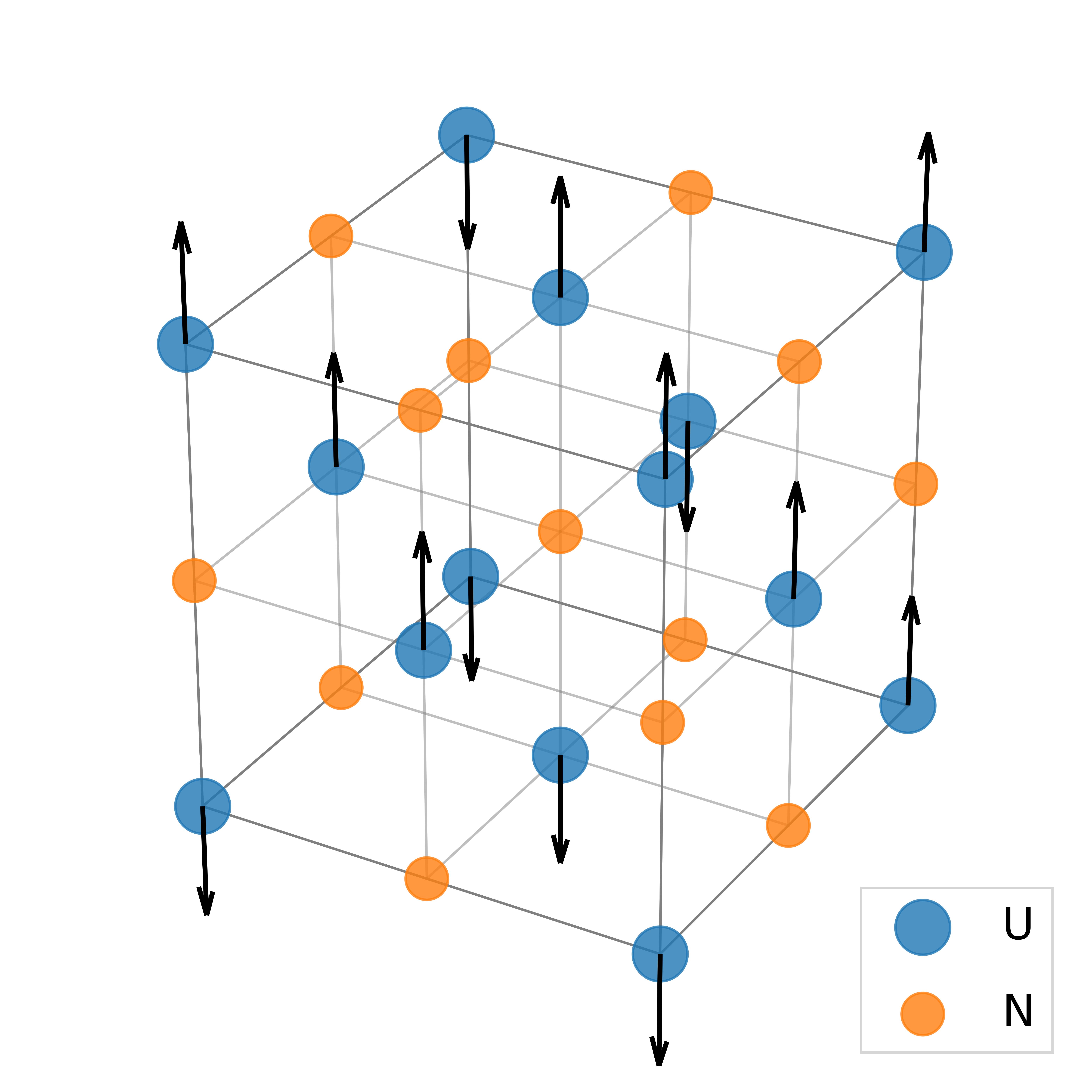}
    \caption{A conventional cubic unit cell of UN with disordered moments assigned to the U atoms. U atoms are blue and N atoms are orange. A single unit cell is shown instead of the used $3 \times 3 \times 3$ supercell for better visibility.}
    \label{Fig:UnitCell}
\end{figure}

UN crystallizes in the NaCl-type structure, with space group symbol $Fm\bar{3}m$ (space group number: 225) \cite{AFLOW}. At room temperature, the lattice parameter is $a$ = 4.89~{\AA}~\cite{Hayes1990I}, and the compound has a melting temperature of 3035~K at a nitrogen vapor pressure of 1~atm~\cite{Hayes1990IV}. In this structure, the uranium atoms occupy the face-centered cubic (FCC) lattice sites, while the nitrogen atoms reside at the octahedral interstitial positions (or vice versa, since the NaCl structure consists of two interpenetrating FCC sublattices). The nearest-neighbor separation between U and N atoms lies along the cube edge and is equal to $a/2$~=~2.445~{\AA}. Owing to the NaCl-type structure, UN exhibits octahedral coordination: each uranium atom is surrounded by six nitrogen atoms located at the vertices of a regular octahedron, and each nitrogen atom is similarly coordinated by six uranium atoms. The conventional unit cell of UN is shown in \cref{Fig:UnitCell}.

Below a Néel temperature, $T_N$, of about 53 K, UN has an antiferromagnetic (AFM) state of type I with the spins of U atoms aligned along the [001] direction \cite{Curry1965}. The expected tetragonal distortion in this type of magnetic order is minimal for UN ($|c/a-1| = 6.5 \times 10^{-4}$ at 4.2 K \cite{Samsel2007}) and it is even debated whether a tetragonal distortion exists or not \cite{Samsel2007}. Above $T_N$, UN transitions to a paramagnetic material and assumes a fully cubic NaCl-type crystal structure.

Paramagnetism is a finite-temperature phenomenon arising from thermal and quantum fluctuations, characterized by random, non-collinear local magnetic moments that persist above the magnetic transition temperature. In this state, these moments fluctuate both in time and space due to temperature-driven spin excitations, resulting in zero time-averaged net magnetization in the absence of an external magnetic field \cite{Abrikosov2016}. The time scale of these fluctuations, known as the spin decoherence time, is on the order of 10 fs \cite{Steneteg2012}, representing the temporal resolution required to correctly simulate spin dynamics in paramagnetic materials. Compared to ordered magnetic states, where excitations tied to spin-wave frequencies occur on a scale of $\sim 100$ fs \cite{Abrikosov2016}, the magnetic degrees of freedom in paramagnetic materials are significantly faster. While the paramagnetic state is ergodic over long timescales, it exhibits transient correlations and non-uniform sampling of the phase space over shorter timescales \cite{Gyorffy1985, Abrikosov2016}. This can be appropriately termed quasi-ergodic or slowly ergodic behavior.

First-principles methods face significant challenges in accurately describing paramagnetism due to both fundamental and practical limitations. Fundamentally, these methods rely on adiabatic decoupling of magnetic and vibrational degrees of freedom \cite{Abrikosov2016} and approximations like the local spin density approximation (LSDA), which can only model ground-state properties and static spin configurations \cite{Giustino2014}. Consequently, they fail to capture dynamic spin fluctuations and strong electron correlations typical of paramagnetic materials \cite{Kotliar2004}. While advanced approaches like DFT+$U$ and the dynamical mean field theory (DMFT) improve correlation modeling, DFT+$U$ cannot represent the paramagnetic state \cite{Liu2019, Shousha2024}, and DMFT struggles with short-range order near the magnetic transition \cite{Abrikosov2016}. Combined methods such as DFT+DMFT offer more accurate descriptions but are computationally intensive \cite{Kotliar2004, Steneteg2012}.

Practically, simulating the high-temperature paramagnetic state with standard \textit{ab initio} molecular dynamics (AIMD) is challenging due to mismatched timescales between atomic and spin dynamics. AIMD typically requires simulation times of 3--5 ps with a time step of 1 fs, while the spin decoherence time is $\sim 10$ fs \cite{Steneteg2012, Abrikosov2016}. As a result, spin configurations evolve more slowly than atomic configurations on short timescales but can change significantly over the duration of an MD simulation. This discrepancy makes it difficult to separate magnetic and vibrational degrees of freedom, leading to the breakdown of adiabatic decoupling.

The disordered local moment (DLM) approach combined with AIMD effectively models magnetic fluctuations and the quasi-static behavior of the paramagnetic state \cite{Abrikosov2016}. This AIMD+DLM framework has been used to study the influence of the paramagnetic state on properties of materials with known strong coupling between magnetic and lattice degrees of freedom. Steneteg \textit{et al.} \cite{Steneteg2012} introduced this method to calculate the equation of state and bulk modulus of CrN, demonstrating no collapse of the bulk modulus across the antiferromagnetic-to-paramagnetic transition. Mozafari \textit{et al.} \cite{Mozafari2016} extended this framework to compute the temperature-dependent elastic constants of CrN. Alling \textit{et al.} \cite{Alling2016} utilized AIMD+DLM to study the impact of lattice vibrations on magnetic and electronic properties of paramagnetic iron.

% Magnetic state of UN at 0 K calculations

Standard 0 K DFT studies of UN using generalized gradient approximation (GGA) exchange-correlation functionals have found that the ferromagnetic (FM) order is lower in energy than AFM UN \cite{Kocevski2022I} in contradiction to experimental observations. In these calculations, FM UN is fully cubic, and AFM UN is characterized by a tetragonal distortion. When an effective Hubbard $U$ = 1.9 eV is used along with the GGA functionals, AFM UN becomes more energetically favorable than FM UN, and AFM UN is characterized by an orthorhombic distortion \cite{Claisse2016b}. Furthermore, Kocevski \textit{et al.} \cite{Kocevski2022I} have shown that when AFM UN is modeled using DFT+$U$, it exhibits imaginary phonon frequencies, indicating a dynamically unstable crystal structure.

This work presents the first application of the AIMD+DLM framework to simulate the paramagnetic state of UN. By treating UN as a paramagnetic material, we calculate several thermophysical properties and compare them with available experimental and AIMD data. To the best of our knowledge, only one AIMD study on UN has been reported in the literature. In that study, Kocevski \textit{et al.} \cite{Kocevski2023} employed FM and AFM models to investigate the thermophysical and elastic properties of UN. However, these models do not accurately represent the high-temperature paramagnetic state of UN, highlighting the significance of our study.

% Therefore, despite being essentially wrong, it appears that FM UN is the most appropriate DFT model of UN at 0 K \cite{Kocevski2022I, AbdulHameed2024d}.

\section{Computational details}

AIMD simulations in this study were performed using the Vienna \textit{ab initio} Simulation Package (VASP) \cite{Kresse1993, Kresse1996a, Kresse1996b}, employing the Perdew-Burke-Ernzerhof (PBE) generalized gradient approximation (GGA) for the exchange-correlation functional \cite{Perdew1996}. Projector-augmented wave (PAW) pseudopotentials were used for uranium and nitrogen. The valence electron configuration of uranium is \( 6s^2 \, 6p^6 \, 6d^2 \, 5f^2 \, 7s^2 \) (14 electrons), while that of nitrogen is \( 2s^2 \, 2p^3 \) (5 electrons). To treat partial electronic occupancies, the first-order smearing method of Methfessel and Paxton \cite{Methfessel1989} was employed with a smearing width of 0.1 eV.

The UN system is modeled using \( 3 \times 3 \times 3 \) supercells containing 216 atoms. A plane-wave cutoff energy of 520 eV is used, and the energy convergence criterion for electronic optimization is set to \( 10^{-3} \) eV. Brillouin-zone integrations are performed at the \( \Gamma \) point. The \( 3 \times 3 \times 3 \) supercells in this work are equilibrated in the \textit{NPT} ensemble at zero pressure and temperatures of 300, 500, 1000, 1500, and 2000 K throughout 1 ps using standard spin-polarized AIMD simulations, in which UN is in the FM state. Subsequently, the disordered local moment (DLM) model is applied.

In the DLM model, it is assumed that the magnetic moments undergo random reorientation over a characteristic spin-flip time, \( t_\mathrm{SF} \) = 5 fs. The spin-flip time is recommended to be much shorter than the spin decoherence time because a slow spin-dynamics model would lead to large spin-lattice correlations if the nuclei were allowed to relax in response to the static local magnetic moment distribution \cite{Steneteg2012}. The simulation is initialized by assigning randomly oriented collinear local magnetic moments to U atoms, under the constraint that the net magnetic moment of the supercell is zero. Collinear spin-polarized AIMD simulations are then performed for \( t_\mathrm{SF} / t_\mathrm{MD} \) time steps, where \( t_\mathrm{MD} \) = 1 fs is the length of a single AIMD time step. After each spin-flip interval, the collinear spins are reassigned randomly while the lattice positions and velocities remain unchanged, and the process is repeated iteratively. During each spin-flip interval, the magnitudes of the local magnetic moments are allowed to evolve self-consistently. The goal is to achieve a magnetic state that exhibits no order on either the supercell length scale or the simulation time scale. Note that the DLM approach alone gives a state of temporarily broken ergodicity, where the system is stuck for a time equal to $t_\mathrm{SF}$ in a single point in the phase space \cite{Gyorffy1985, Abrikosov2016}. However, the occasional spin flips motivated by the self-consistent solution make the system achieve the desired quasi-ergodic behavior of the paramagnetic state. Spin-orbit coupling (SOC), i.e., the interaction between the electron spins and the atomic orbits in magnetic atoms \cite{Giustino2014}, as well as its key ingredient, i.e., non-collinear magnetic moments, are not treated in this work. The DLM iterations are continued for 3 ps, which corresponds to 600 spin flips. In total, the 1-ps FM run and the 3-ps DLM run give a total simulation time of 4 ps. A Bash/Python script that implements the DLM method using VASP can be found in the Supplementary Information, and its workflow is detailed in \cref{AlgoX} in the appendix.

To demonstrate the accuracy of the AIMD+DLM model of UN, its lattice constant and specific heat capacity are calculated and compared to available experimental data. The lattice constant is calculated by averaging the supercell dimensions along the $x$-, $y$- and $z$-directions over the whole simulation time. UN behaves as a metallic solid \cite{Yang2021}, and its specific heat capacity is a sum of lattice and electronic specific heats: $C_P = C_\mathrm{lat} + C_\mathrm{elec}$. The total energies of supercells are averaged over the last 2.5 ps for all temperatures. Then, the averaged values are fitted to a third-degree polynomial of the form:
\begin{equation}
E = a + b T + c T^2 + d T^3.
\end{equation}
The lattice specific heat capacity can then be calculated as:
\begin{equation}
C_\mathrm{lat} = \frac{1}{n} \frac{dE}{dT} = \frac{1}{n} \left( b + 2 c T + 3 d T^2 \right),
\end{equation}
where $n$ is the number of moles. This method was employed instead of a finite difference approach due to the limited number of temperature points. The electronic specific heat is estimated from \cite{Gopal1966}:
\begin{equation}
C_\mathrm{elec} = \gamma T,
\end{equation}
where $\gamma$ is the electronic specific heat coefficient (J/mol-K$^2$), whose value is $\gamma$ = 3.7 mJ/mol-K$^2$ \cite{Samsel2007,AbdulHameed2024}.

\section{Results}

When the initial $3 \times 3 \times 3$ UN supercell is constructed with an AFM configuration, the system transitions to an FM state almost instantaneously upon the onset of equilibration. Tests with the smaller $2 \times 2 \times 2$ AFM supercell show that it can survive for about 1~ps at 300~K, before finally collapsing to the FM state. For the larger $3 \times 3 \times 3$ supercell, however, the AFM state could not be sustained at any temperature. This occurs when the spin configuration is allowed to stabilize to its lowest energy configuration. On the other hand, when the AFM state is enforced, by setting \verb|NUPDOWN| = 0 \cite{NUPDOWN}, the AFM cannot be sustained. However, expectedly, the AFM supercell has higher energy than that of the FM supercell. It should be noted that when the AFM state is enforced, the supercell remains cubic, consistent with the results of Kocevski \textit{et al.} \cite{Kocevski2022I}. This behavior is likely due to the small magnitude of the zero-temperature tetragonal distortion, which is effectively suppressed by thermal fluctuations at finite temperatures.

\begin{figure}[h!]
\centering
\begin{subfigure}{0.48\textwidth}
    \includegraphics[width=\textwidth]{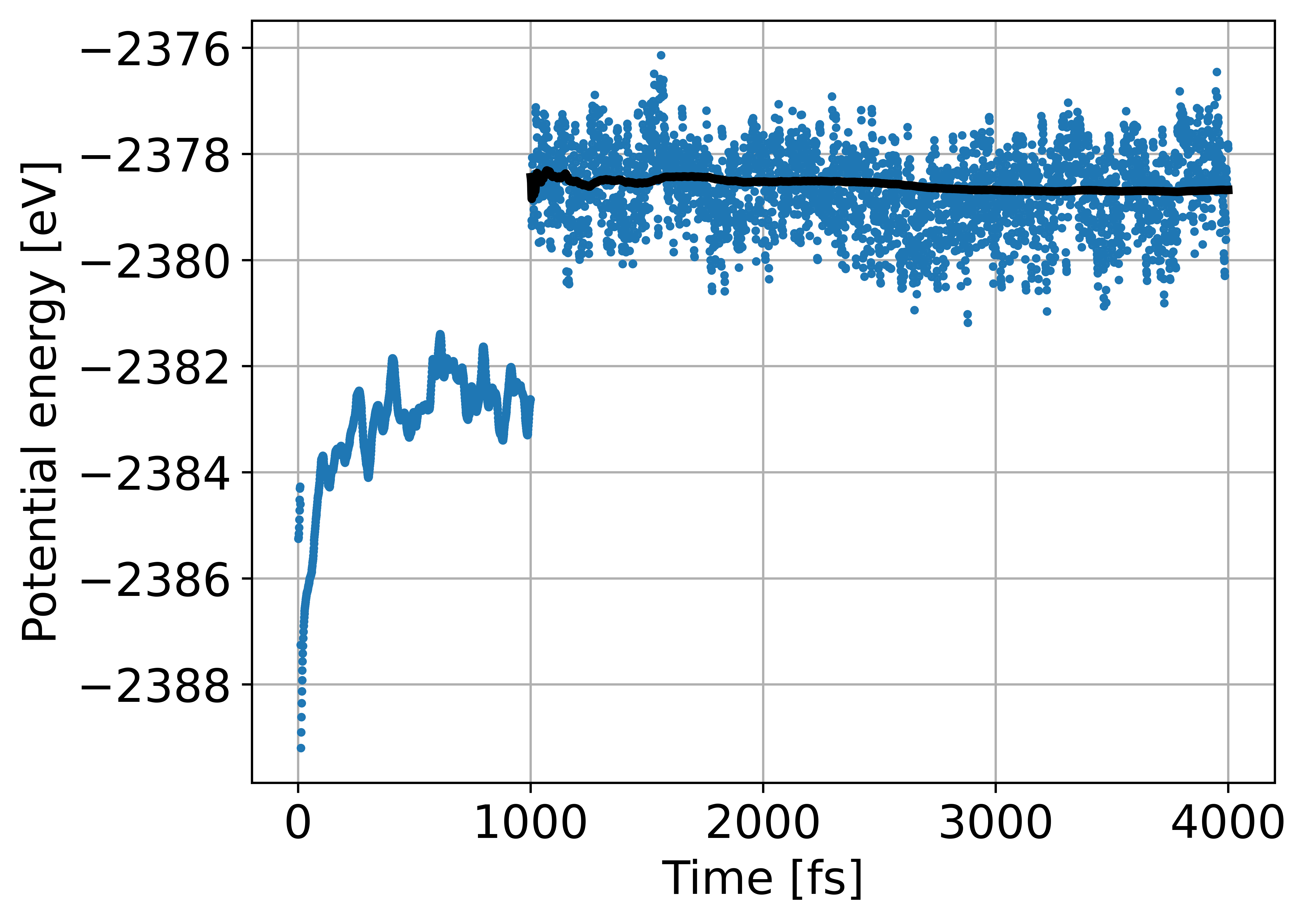}
    \caption{$T$ = 300 K}
    \label{Fig:300-PE}
\end{subfigure}
\hfill
\begin{subfigure}{0.48\textwidth}
    \includegraphics[width=\textwidth]{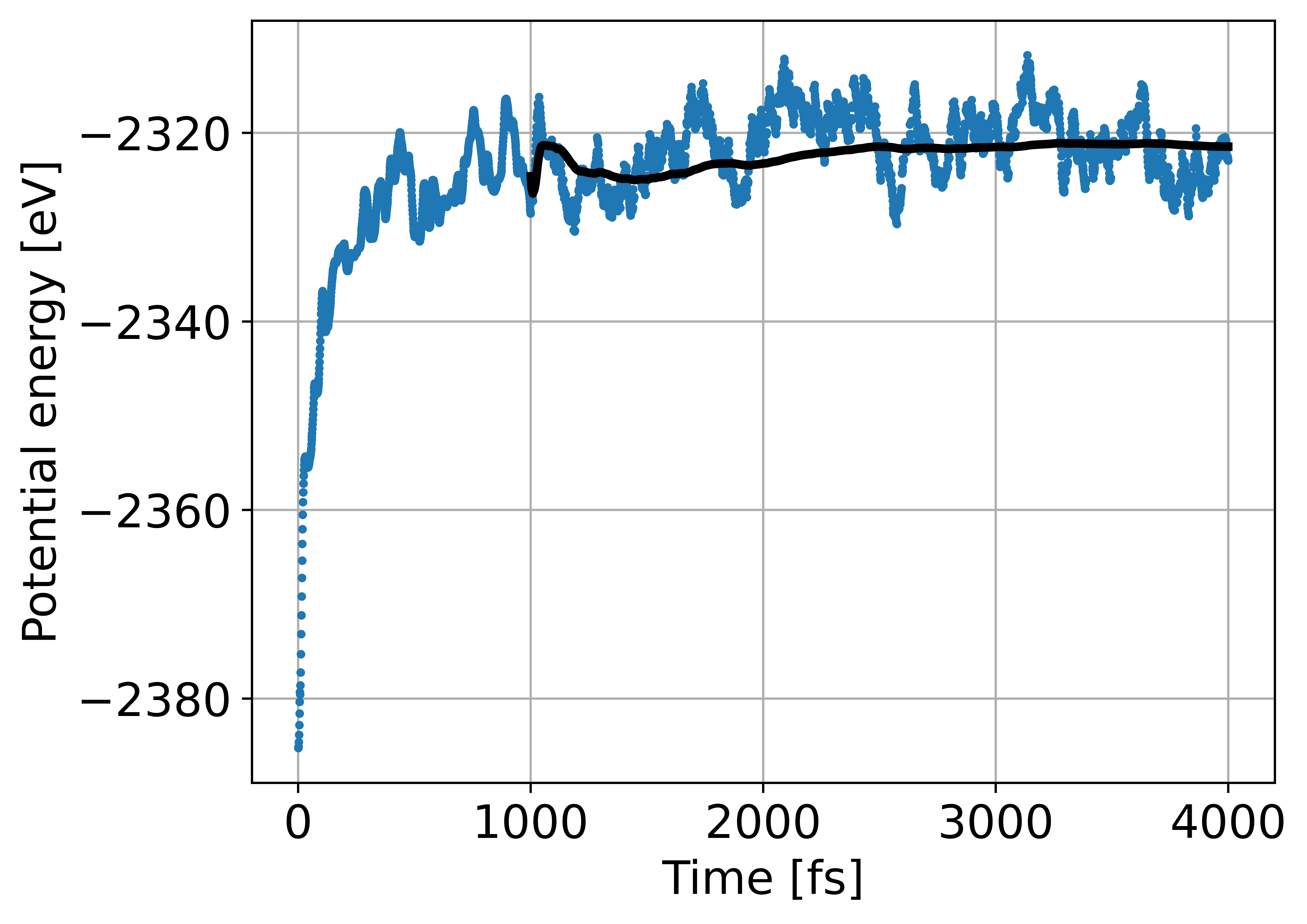}
    \caption{$T$ = 2000 K}
    \label{Fig:2000-PE}
\end{subfigure}
\caption{(Color online) Potential energy of the supercell at \textbf{(a)} 300 K and \textbf{(b)} 2000 K. Black thick lines correspond to running averages.}
\label{fig:PE}
\end{figure}

The total system potential energy of UN as a function of time at 300 and 2000 K are shown in \cref{fig:PE}. In \cref{Fig:300-PE}, it can be seen that for low temperatures, the potential energy of the FM state (the first 1 ps in the figure) is lower than that of the DLM state (the last 3 ps in the figure). With increasing temperature, e.g., \cref{Fig:2000-PE}, the potential energies of the two states become comparable. Despite the large thermal fluctuations in the potential energy at 2000 K, its running average is seen to be well-converged during the last 1.5 ps of the simulation time.

\begin{figure}[h!]
\centering
\begin{subfigure}{0.48\textwidth}
    \includegraphics[width=\textwidth]{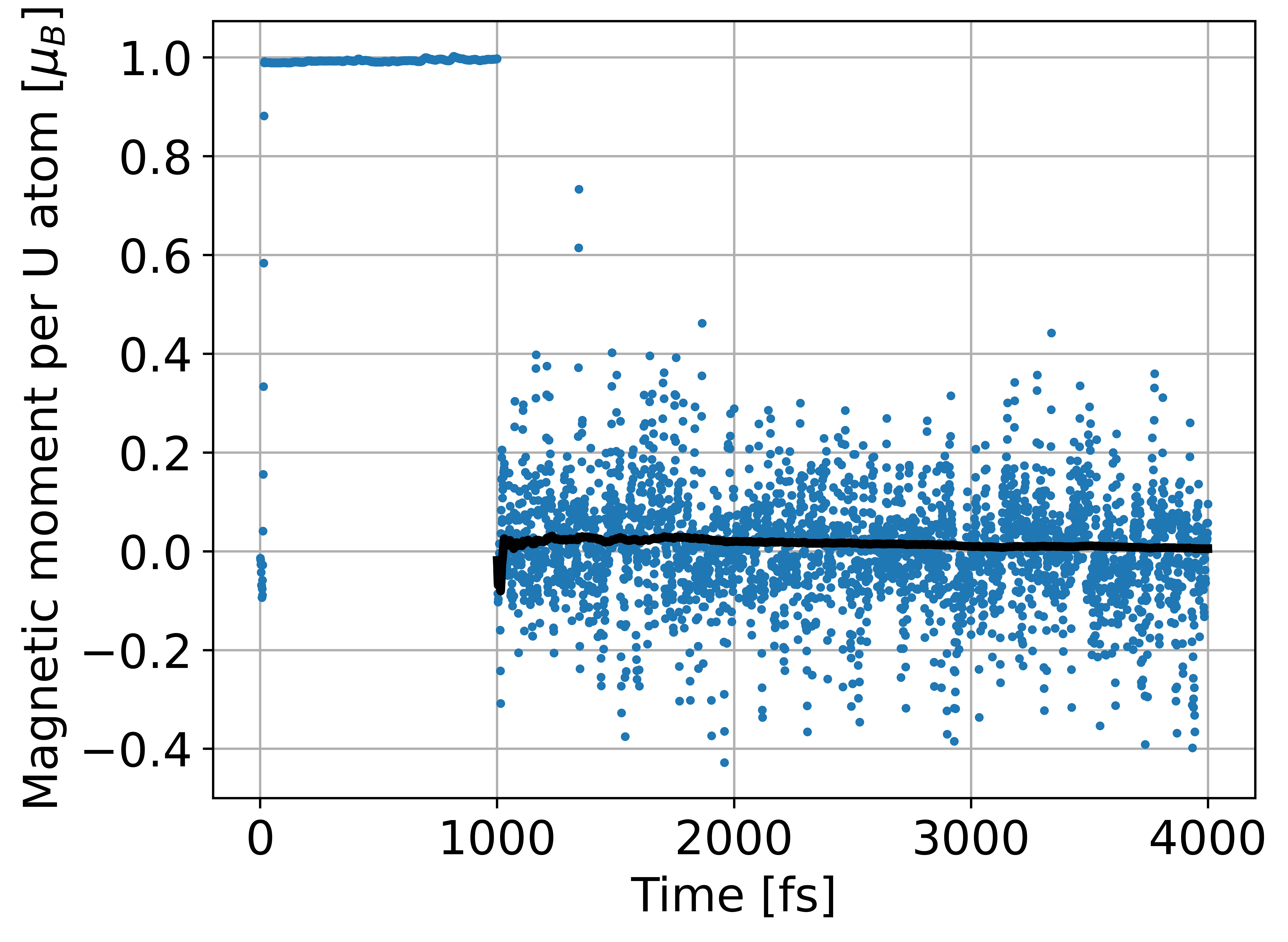}
    \caption{$T$ = 500 K}
    \label{Fig:500-MagMom}
\end{subfigure}
\hfill
\begin{subfigure}{0.48\textwidth}
    \includegraphics[width=\textwidth]{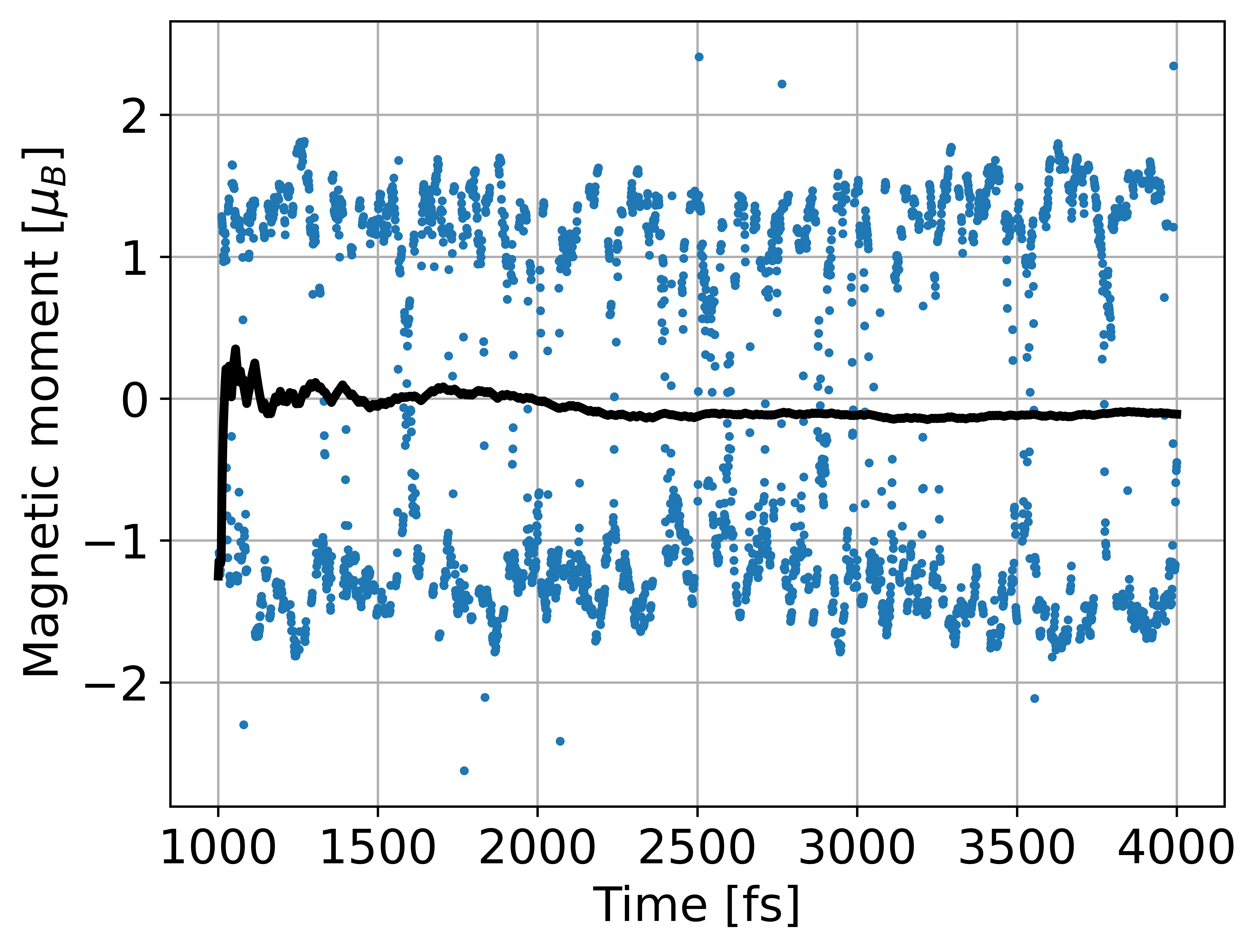}
    \caption{$T$ = 1000 K}
    \label{Fig:AtomMagMom-1000}
\end{subfigure}
\caption{(Color online) \textbf{(a)} Magnetic moment per U atom at 500 K averaged over the supercell. \textbf{(b)} Magnetic moment of an arbitrarily chosen U atom at 1000 K for only the DLM part of the simulation. Black thick lines correspond to running averages.}
\label{fig:magmom}
\end{figure}

\Cref{fig:magmom} shows the system-wide magnetic moment and the magnetic moment of an arbitrarily chosen U atom in the supercell as a function of time. In \cref{Fig:500-MagMom}, it can be seen that the average magnetic moment per U atom at 500 K is practically conserved and fluctuates about zero during the whole time of the DLM run, where the fluctuations are bound between $\pm 0.25 \ \mu_B$. The magnetic moment of an arbitrarily chosen U atom at 1000 K is shown in \cref{Fig:AtomMagMom-1000}. It fluctuates about a running average of zero where the fluctuations are mostly between $\pm 1.25 \ \mu_B$, which coincides with the per-U magnetic moment of FM UN at 0 K \cite{Kocevski2022I}, but is larger than $\mu = 0.75 \ \mu_B$---the experimental per-U magnetic moment of AFM UN below $T_N$ \cite{Curry1965}. Similar trends have been observed at all considered temperatures.

It should be noted that these fluctuations represent the instantaneous spin-only local moments, calculated self-consistently. As shown by Gryaznov \textit{et al.}~\cite{Gryaznov2012} for UN and Shorikov \textit{et al.}~\cite{Shorikov2005} for PuN, inclusion of SOC generates an orbital moment that is anti-parallel to the spin moment, resulting in a reduced total magnetic moment ($\sim 0.8 \mu_B$) closer to the experimental value. However, this reduction arises from the addition of a negative orbital component, not from a smaller spin moment. The spin-only moment of $\sim 1.25 \mu_B$ predicted in our DLM simulations is therefore close to the correct intrinsic spin moment, consistent with prior DFT results in the absence of SOC. Our AIMD+DLM framework accurately captures this value and its thermally driven fluctuations, offering both a qualitatively and quantitatively correct description of the high-temperature magnetic state---something that static, ordered-moment approaches cannot achieve. % It should be noted that the magnitudes of the spin magnetic moment and the antiparallel orbital magnetic moment are sensitive to the choice of exchange-correlation functional and to whether or not the Hubbard $U$ correction is included.

To converge to the magnetic ground state, the VASP documentation recommends setting the initial magnetic moments larger than the expected values \cite{MAGMOM}. In our calculations, we set the initial magnetic moments per U atom to $\pm 5 \ \mu_B$. To determine if initial magnetic moments affect spin moment oscillations, we performed AIMD+DLM calculations at 300 K for 0.5 ps following a 1-ps FM equilibration with initial magnetic moments per U atom set to $\pm 1 \ \mu_B$ instead of $\pm 5 \ \mu_B$. The per-atom magnetic moments fluctuated between $\pm 1.25 \ \mu_B$, indicating that the prescribed magnitude of the initial magnetic moments does not influence the magnitude of the observed fluctuations. That is, the steady-state fluctuations expected of a successful AIMD+DLM simulation are robust to the initial conditions.

\begin{figure}[h!]
\centering
\begin{subfigure}{0.48\textwidth}
    \includegraphics[width=\textwidth]{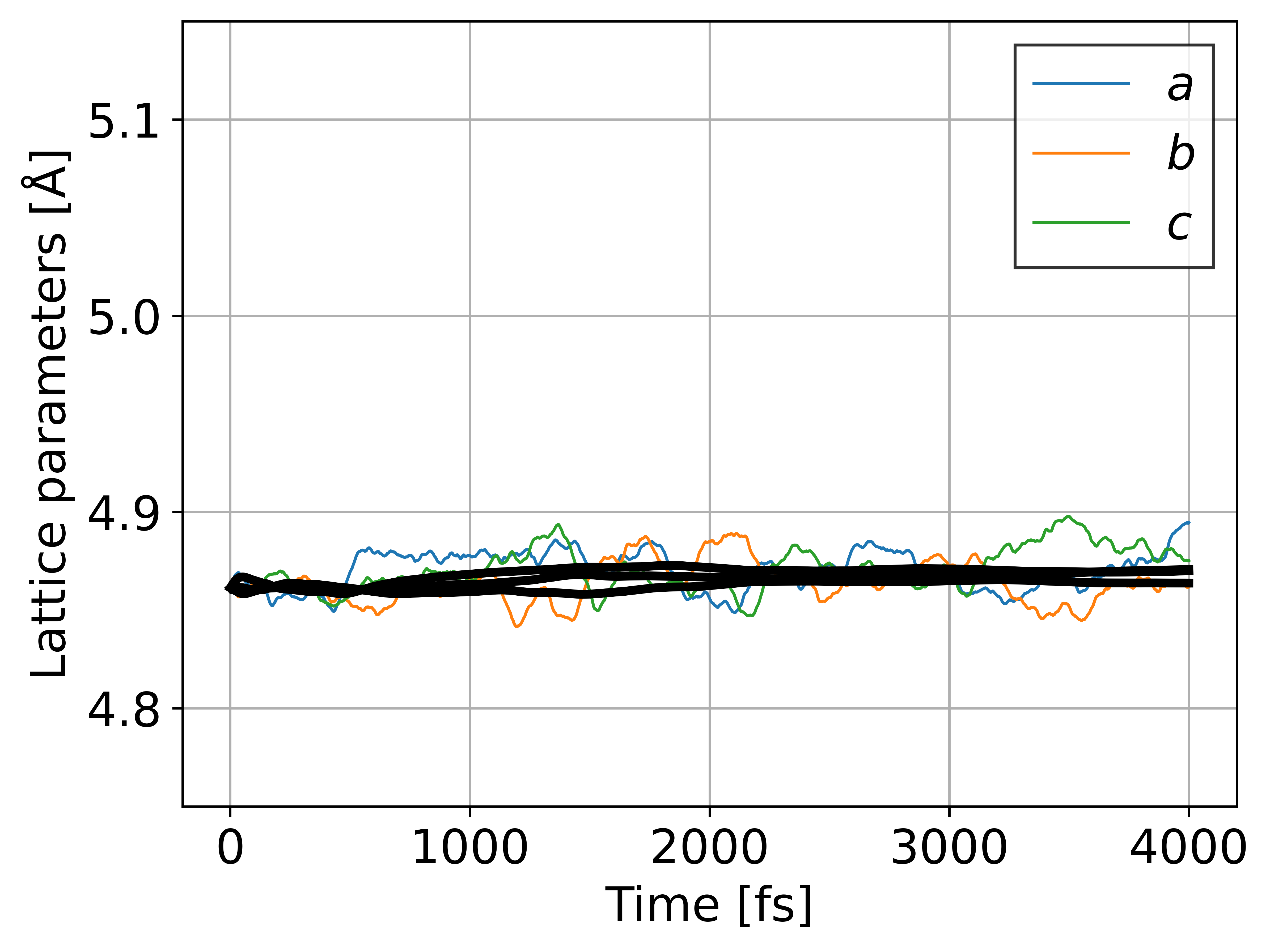}
    \caption{$T$ = 300 K}
    \label{Fig:LatConst-300}
\end{subfigure}
\hfill
\begin{subfigure}{0.48\textwidth}
    \includegraphics[width=\textwidth]{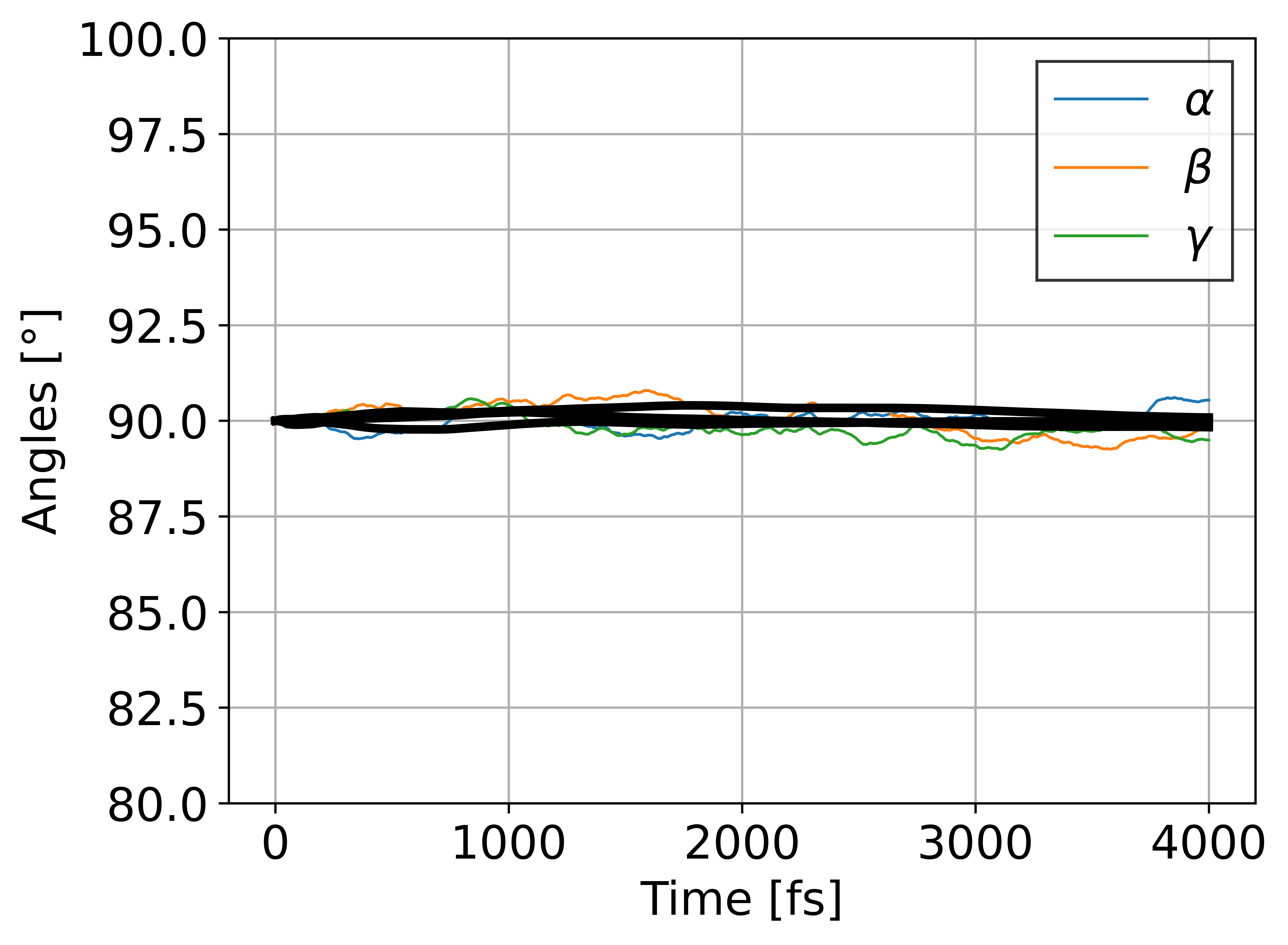}
    \caption{$T$ = 300 K}
    \label{Fig:Angles-300}
\end{subfigure}
\begin{subfigure}{0.48\textwidth}
    \includegraphics[width=\textwidth]{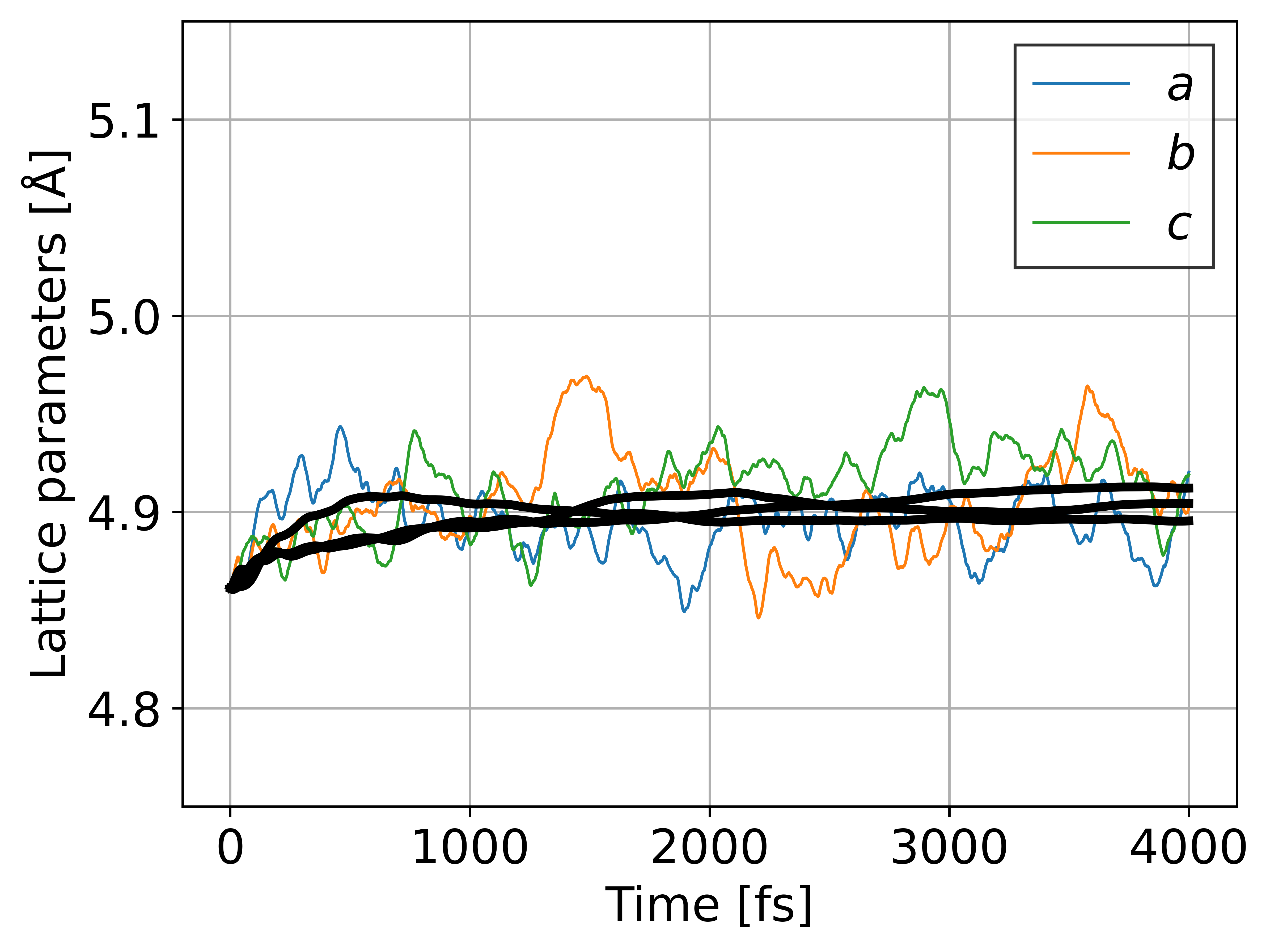}
    \caption{$T$ = 1500 K}
    \label{Fig:LatConst-1500}
\end{subfigure}
\hfill
\begin{subfigure}{0.48\textwidth}
    \includegraphics[width=\textwidth]{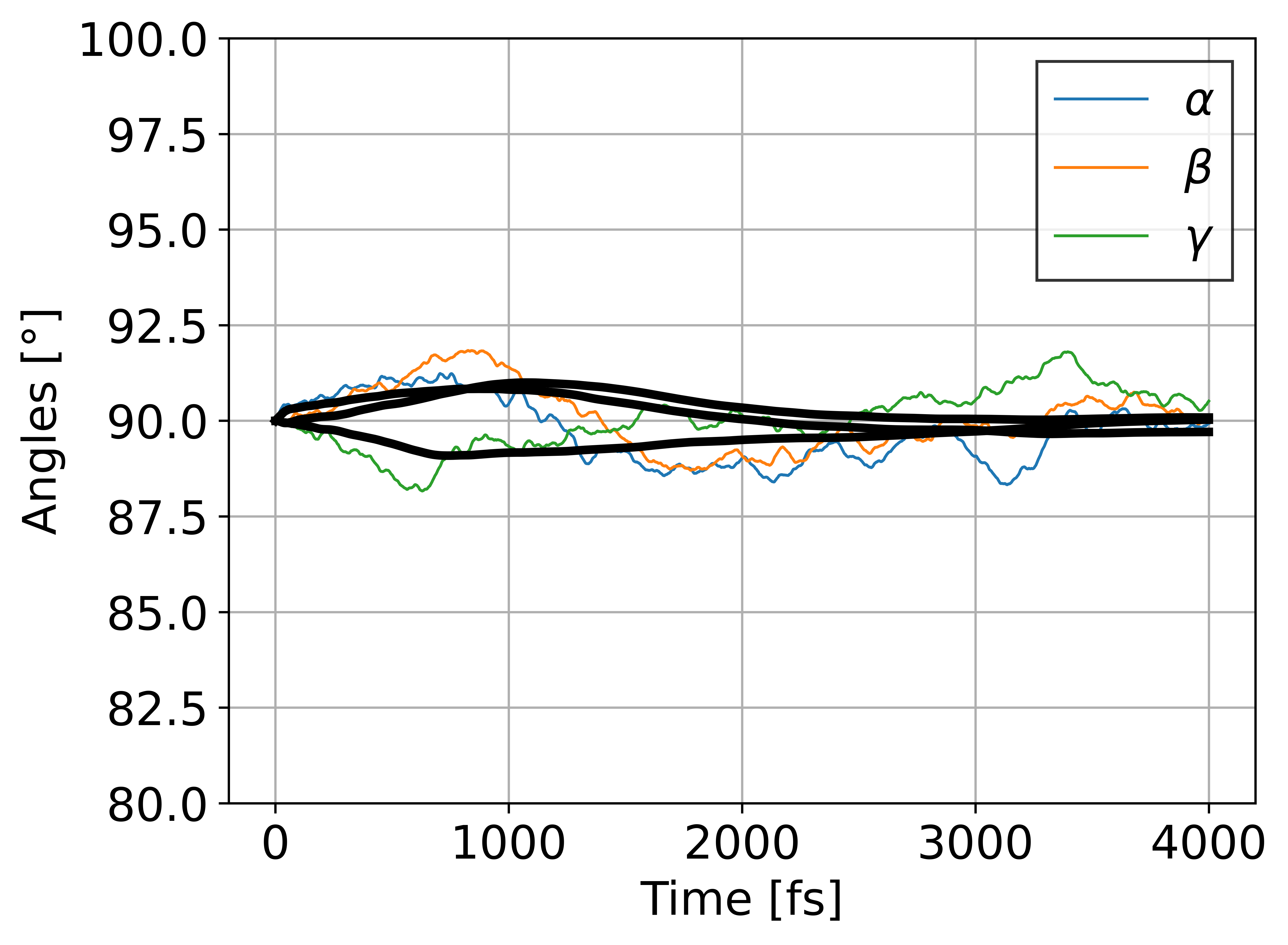}
    \caption{$T$ = 1500 K}
    \label{Fig:Angles-1500}
\end{subfigure}
\caption{(Color online) \textbf{(a)} Lattice parameters ($a$ : blue; $b$ : orange; $c$ : green)  of the UN unit cell at 300 K. \textbf{(b)} Angles ($\alpha$ : blue; $\beta$ : orange; $\gamma$ : green) between the lattice vectors of the UN unit cell at 300 K. \textbf{(c)} Lattice parameters of the UN unit cell at 1500 K. \textbf{(d)} Angles between the lattice vectors of the UN unit cell at 1500 K. Black thick lines correspond to running averages.}
\label{Fig:struct}
\end{figure}

The lattice parameter and structure angles are shown in \cref{Fig:struct} as a function of time at 300 and 1500 K. 
The running averages of the UN unit cell's lengths and angles nearly coincide at lower temperatures as is obvious in \cref{Fig:LatConst-300,Fig:Angles-300}, respectively. As expected, deviations from the purely cubic structure increase with increasing temperature due to thermal fluctuations, but the deviations are relatively small and converge to the purely cubic case when the DLM run is well-converged, as can be seen in \cref{Fig:LatConst-1500,Fig:Angles-1500} for $T$ = 1500 K. Thus, it can be concluded that the DLM model of paramagnetic UN predicts it to be purely cubic at all temperatures, in agreement with experimental observations. % The reason why the DLM gives a cubic structure and not the tetragonal distortion expected from an AFM order can be explained as follows: The spin-flip time interval, $t_\mathrm{SF}$ = 5 fs, is so short that the nuclei do not have sufficient time to adjust their positions for the current distribution of magnetic moments and move toward the positions that would give a tetragonal/orthorhombic distortion \cite{Steneteg2012}.

\begin{figure}[h!]
\centering
\begin{subfigure}{0.48\textwidth}
    \includegraphics[width=\textwidth]{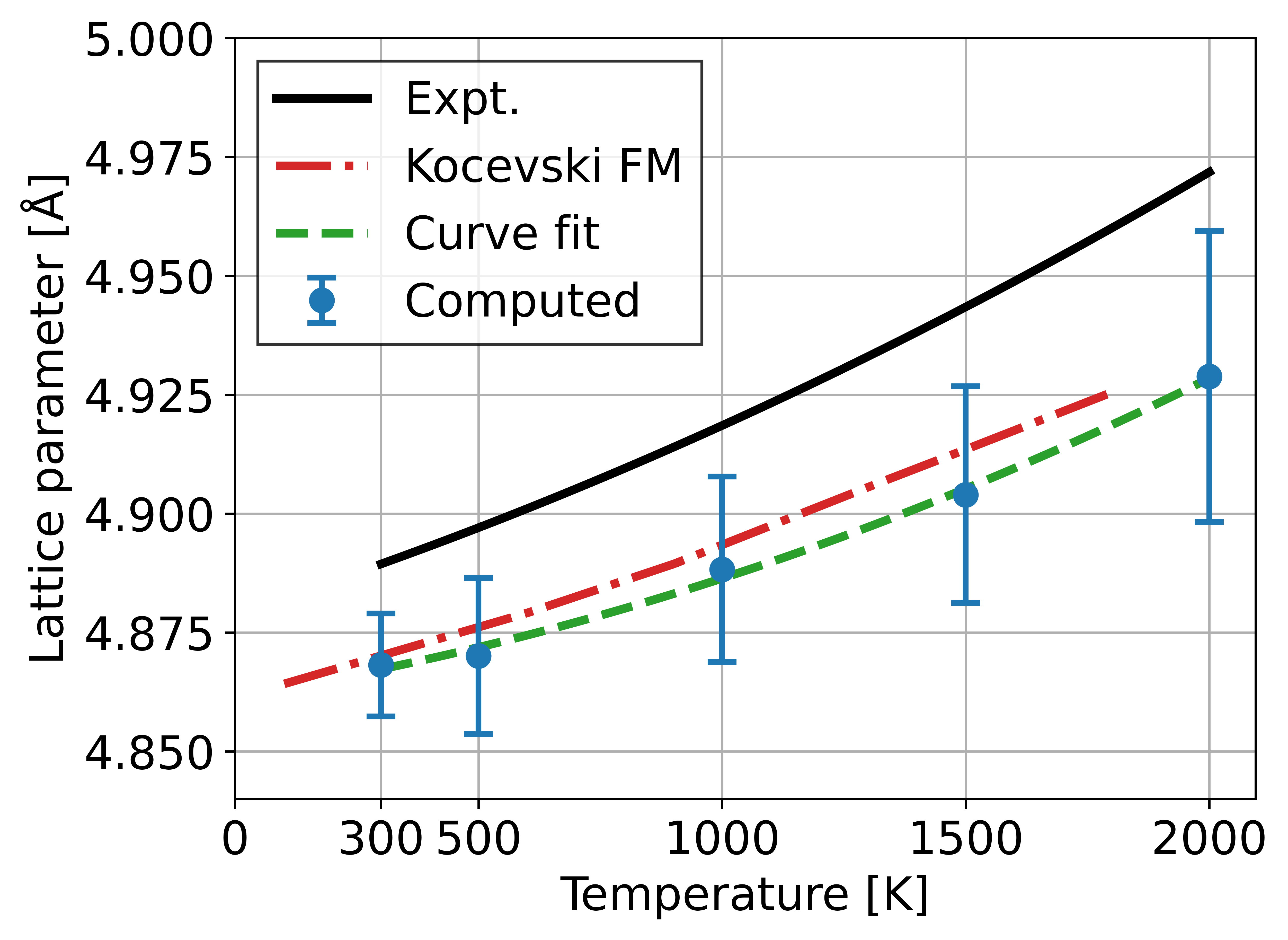}
    \caption{}
    \label{Fig:LatConst}
\end{subfigure}
\hfill
\begin{subfigure}{0.48\textwidth}
    \includegraphics[width=\textwidth]{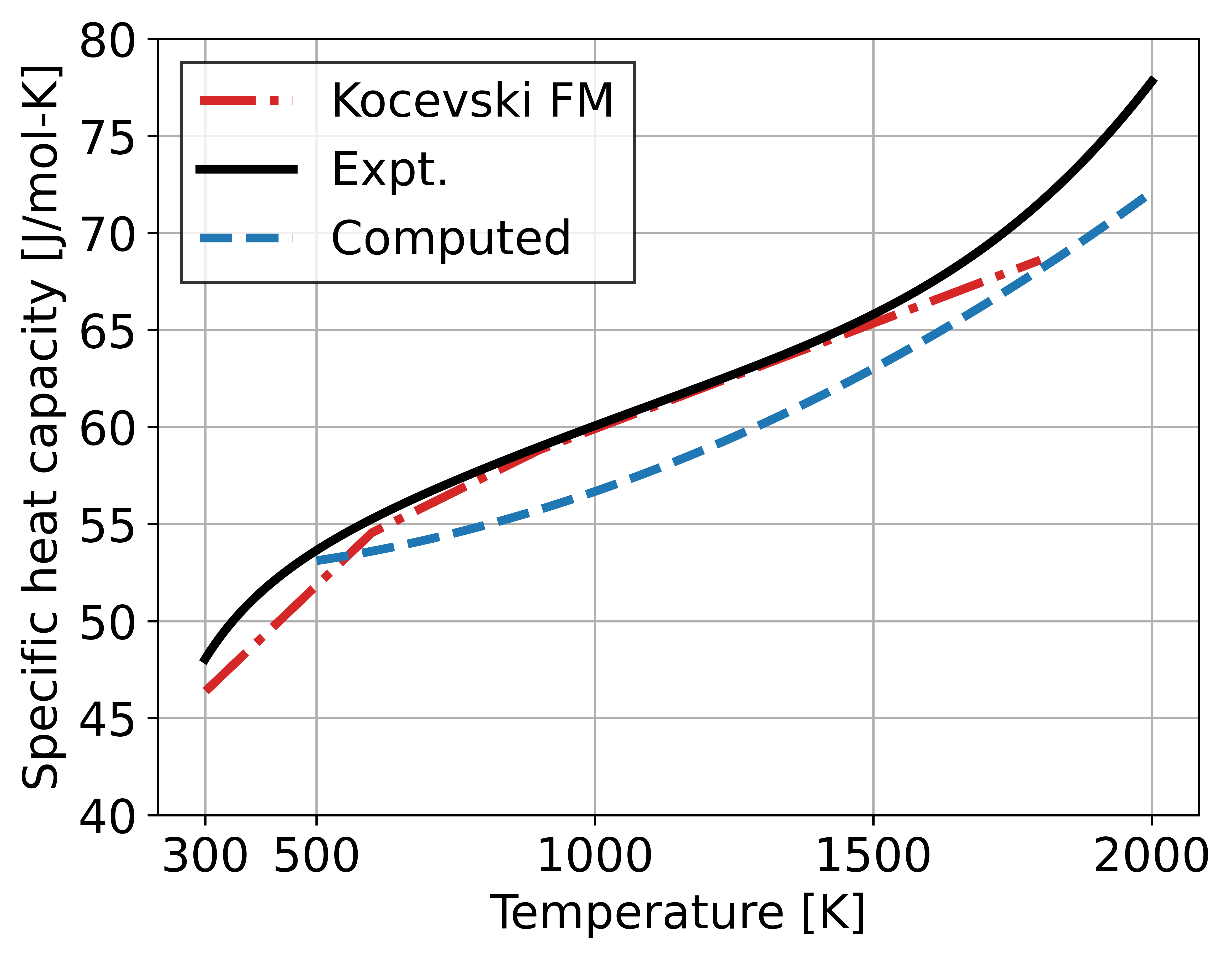}
    \caption{}
    \label{Fig:CP}
\end{subfigure}
\caption{(Color online) \textbf{(a)} Calculated lattice constant of UN compared to the empirical correlation of Hayes \textit{et al.} \cite{Hayes1990I} and the AIMD data calculated by Kocevski \textit{et al.} \cite{Kocevski2023} for FM UN. Error bars correspond to one standard deviation of the time average. \textbf{(b)} Fitted specific heat capacity of UN compared to the empirical correlation of Hayes \textit{et al.} \cite{Hayes1990IV} and the AIMD data calculated by Kocevski \textit{et al.} for FM UN.}
% \label{}
\end{figure}

The lattice parameter of UN obtained from AIMD simulations is presented in \cref{Fig:LatConst}. The calculated lattice parameter is somewhat underestimated compared to the empirical correlation reported by Hayes \textit{et al.} \cite{Hayes1990I}, and is very close to that calculated by Kocevski \textit{et al.} \cite{Kocevski2023} for FM UN using AIMD. The standard deviation of the lattice parameter exhibits a systematic increase with temperature, attributed to enhanced thermal fluctuations. The temperature dependence of the lattice parameter has been fitted to the following polynomial expression:
\begin{equation}
a(T) = 4.862 + 1.593 \times 10^{-5} T + 8.719 \times 10^{-9} T^2,    
\end{equation}
with a coefficient of determination, \(R^2 = 99.6\%\).

The calculated specific heat capacity, $C_P$, of UN is presented in \cref{Fig:CP}, alongside a comparison with the empirical correlation provided by Hayes \textit{et al.} \cite{Hayes1990IV} and the AIMD data calculated by Kocevski \textit{et al.} \cite{Kocevski2023} for FM UN. At $T > 500$ K, while \( C_P \) is slightly underestimated, it closely follows the same trend as the empirical correlation. This is not the case for Kocevski \textit{et al.}'s calculation, which follows a linear increase that would underestimate $C_P$ if extrapolated to higher $T$.

\section{Discussion}

While the AIMD study of Kocevski \textit{et al.} \cite{Kocevski2023} using FM and AFM models yielded reasonable predictions for certain thermophysical properties, they do not faithfully capture the disordered magnetic character of the paramagnetic phase. The AIMD+DLM method addresses this shortcoming by explicitly incorporating magnetic disorder and its dynamic interaction with the lattice. The specific heat capacity, $C_P$, calculated by Kocevski \textit{et al.} exhibits an asymptotically linear increase with temperature. Our calculated $C_P$, while slightly underestimated, displays a power-law increase in closer agreement with the experimental trend despite using fewer temperature points.

An important extension of this work would be the inclusion of SOC within the AIMD+DLM framework. SOC plays a significant role in actinide compounds like UN by contributing orbital components to the total magnetic moment. Incorporating SOC would enhance the physical realism of the simulations and improve both the quantitative accuracy and qualitative fidelity of the modeled magnetic state. Although computationally more demanding, AIMD+DLM+SOC represents a natural next step toward more comprehensive modeling of magnetic-lattice coupling in $f$-electron systems.

The central purpose of this work is twofold: To demonstrate that the AIMD+DLM methodology can be systematically applied to simulate the paramagnetic state of actinide compounds and to provide a modular, automated workflow for its implementation in VASP. The accompanying Bash/Python script offers a practical, user-friendly tool to streamline AIMD+DLM simulations and supports reproducibility and scalability in future computational studies.

Beyond the scope of thermophysical properties, this methodology enables the investigation of magnetic-lattice coupled phenomena such as thermal conductivity and defect energetics under realistic high-temperature conditions. Its applicability to $f$-electron materials makes AIMD+DLM a valuable asset for studying a wide class of strongly correlated magnetic materials, including advanced nuclear fuels.

\section{Conclusions}

This study presents a systematic application of \textit{ab initio} molecular dynamics combined with the disordered local moments model to investigate the paramagnetic state of UN. The main conclusions are as follows:

\begin{itemize}
    \item The AIMD+DLM framework effectively captures the paramagnetic nature of UN, modeling quasi-static magnetic fluctuations and their coupling to lattice vibrations, and thus overcomes key limitations of standard first-principles approaches.
    
    \item The method predicts a purely cubic structure across the investigated temperature range, in agreement with experimental observations of the paramagnetic phase of UN.
    
    \item The lattice parameter is slightly underestimated relative to the empirical fit by Hayes \textit{et al.}, consistent with previous FM/AFM computational models.
    
    \item The calculated specific heat capacity closely follows experimental trends. Unlike the asymptotically linear increase predicted by prior AIMD studies, our results exhibit a power-law increase with temperature, in better agreement with measurements despite the use of fewer data points.
    
    \item A secondary but significant contribution of this work is the development of a general-purpose Bash/Python script that automates AIMD+DLM simulations in VASP. This tool facilitates systematic, extensible, and reproducible modeling of paramagnetic materials for both elemental and compound systems.
\end{itemize}

In summary, the AIMD+DLM method is a robust and transferable tool for simulating high-temperature magnetic states and associated thermophysical behaviors in actinide materials. Its systematic implementation broadens the potential for high-fidelity simulations in nuclear materials research and beyond.

\section{Acknowledgments}

This research made use of the resources of the High-Performance Computing Center at Idaho National Laboratory, which is supported by the Office of Nuclear Energy of the U.S. Department of Energy and the Nuclear Science User Facilities under Contract No. DE-AC07-05ID14517.

\newpage
\section{Appendix}

\begin{algorithm}
\caption{Workflow for Automated AIMD+DLM Simulations in VASP}
\begin{algorithmic}[1]
\Statex \textbf{Initial Equilibration Run}
\State Set \texttt{MAGMOM} with zero-spin atoms first
\State Order atoms accordingly in \texttt{POSCAR}
\State Run FM simulation to reach equilibrium (Run 0)

\vspace{1ex}
\Statex \textbf{Prepare for DLM Simulation}
\State Edit \texttt{DLMSequence.sh} to match HPC job system
\State In folder \texttt{0}, include input/output files from Run 0, and a short-run job submission script (e.g., \texttt{pbs.sh})
\State In parent folder, place: \texttt{DLM.in}, \texttt{ModifyINCAR.py}, \texttt{DLMSequence.sh}
\State Edit \texttt{DLM.in} with parameters: \texttt{num\_spin\_atoms}, \texttt{tot\_num\_atoms}, \texttt{spin\_flip\_time}, \texttt{time\_step}, and \texttt{simulation\_time}
\State Execute the sequence with \texttt{bash ./DLMSequence.sh}

\vspace{1ex}
\Statex \textbf{Spin Flipping Loop Inside \texttt{DLMSequence.sh}}
\For{$j = 1$ to $\texttt{simulation\_time} / (\texttt{spin\_flip\_time} / \texttt{time\_step})$}
    \State Wait for completion of Run $j-1$
    \State Copy files from Run $j-1$ into folder $j$
    \State Remove spin history by trimming magnetization section from \texttt{CHGCAR} 
    \State Call \texttt{ModifyINCAR.py} to set random \texttt{MAGMOM} values and \texttt{NSW} for each run
    \State Replace \texttt{INCAR} with modified version
    \State Submit job via HPC scheduler (e.g., \texttt{qsub pbs.sh})
\EndFor
\end{algorithmic}
\label{AlgoX}
\end{algorithm}

\FloatBarrier

\bibliographystyle{elsarticle-num}
\bibliography{ref}

\end{document}